\documentclass{article}

\usepackage{arxiv}

\usepackage[utf8]{inputenc} 
\usepackage[T1]{fontenc}    
\usepackage{hyperref}       
\usepackage{url}            
\usepackage{booktabs}       
\usepackage{amsfonts}       
\usepackage{nicefrac}       
\usepackage{microtype}      
\usepackage{graphicx}
\usepackage[numbers]{natbib}
\usepackage{doi}
\usepackage{xcolor}
\usepackage[version=4]{mhchem}
\usepackage[english]{babel}
\usepackage{microtype}
\usepackage{amsmath,amssymb,amsfonts}
\usepackage{algorithmic}
\usepackage{comment}
\usepackage{textcomp}
\usepackage{listings}

\begin{document}

\title{Iterative Clustering Material Decomposition Aided by Empirical Spectral Correction for High Resolution Photon Counting Detectors in Micro CT }
\author{Juan C R Luna and Mini Das*\\
Department of Physics, University of Houston, Houston 77204\\
*mdas@uh.edu}





\maketitle

\begin{abstract}
Photon counting detectors (PCDs) offer promising advancements in computed tomography (CT) imaging by enabling the quantification and 3D imaging of contrast agents and tissue types through multi-energy projections. However, the accuracy of these decomposition methods hinges on precise composite spectral attenuation values that one must reconstruct from spectral micro CT. Factors such as surface defects, local temperature, signal amplification, and impurity levels can cause variations in detector efficiency between pixels, leading to significant quantitative errors. In addition, some inaccuracies such as the charge sharing effects in PCDs are amplified with a high Z sensor material and also with a smaller detector pixels that are preferred for micro CT. In this work, we propose a comprehensive approach that combines practical instrumentation and measurement strategies leading to quantitation of multiple materials within an object in a spectral micro CT with a photon counting detector. Our Iterative Clustering Material Decomposition (ICMD) includes an empirical method for detector spectral response corrections, cluster analysis and multi-step iterative material decomposition. Utilizing a CdTe-1mm Medipix detector with a 55$\mu m$ pitch, we demonstrate the quantitatively accurate decomposition of several materials in a phantom study, where the sample includes mixtures of material, soft material and K-edge materials. We also show an example of biological sample imaging and separating three distinct types of tissue in mouse: muscle, fat and bone. Our experimental results show that the combination of spectral correction and high-dimensional data clustering enhances decomposition accuracy and reduces noise in micro CT. This ICMD allows for quantitative separation of more than three materials including mixtures and also effectively separates multi-contrast agents.
\end{abstract}


\section{Introduction}
\label{sec:introduction}
Photon counting detectors (PCDs) offer a significant advancement in computed tomography (CT) systems by allowing the separation of incoming photons into multiple energy bins, resulting in fast, single-exposure multi-energy CT images. This spectrally resolved information leads to spectral X-ray tomography and material decomposition, with potential applications in biomedical imaging, food and agricultural sciences, security inspection, and industrial radiography. For instance, multi-energy tomography enables imaging and decomposition of a wide range of contrast agents for biomedical applications~\cite{tao2019feasibility, symons2017photon, fredette2019multi}.

Despite the promise of PCDs to obtain spectral information in x-ray measurements, they are an emerging technology and suffer from spectral distortions due to various physical effects. Some examples are charge sharing, fluorescence, K-escape energy loss, pulse pileup, and inter-pixel variability~\cite{ren2018tutorial, taguchi2013vision}. Such distortions result in inaccurate quantitation of spectral X-ray attenuation and lead to noise and ring artifacts in CT reconstructions~\cite{jakubek2007data}. Therefore, spectral distortion correction is essential for enhancing the quantitative accuracy in spectral CT and material decomposition performance.

\begin{figure}[!h]
\centering
\includegraphics[width=0.5\textwidth]{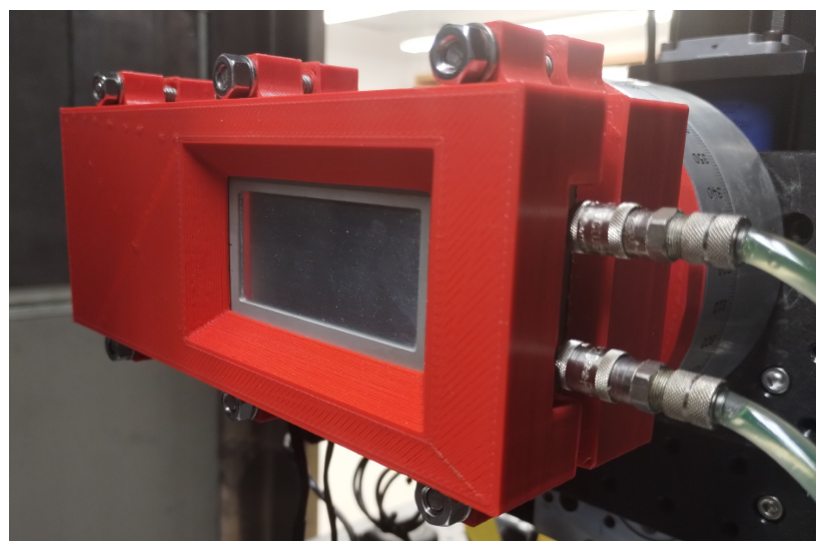}
\caption{\textcolor{black}{Our data was acquired using the WidePix CdTe detector, a tiled assembly of five Medipix3RX single chip units, in CSM mode. The detector features 1000 $\mu m$ CdTe pixels with a 55 $\mu m$ pitch, providing high spatial resolution. Its active area measures 1.408 cm $\times$ 7.040 cm.}}
\label{fig:Widepix}
\end{figure}

Several spectral correction methods have been proposed, yet many of these approaches require a detailed formulation of the detector's response function and often fail to account for pixel-to-pixel variations in individual detectors. Alternative approaches, such as artificial neural networks (ANN), demand complex training processes, prior knowledge of ground truth and distortion-free projections~\cite{touch2016neural}.

Even with ideal spectral correction, an accurate multi-material decomposition is inherently challenging due to the ill-conditioned nature of the problem. Many existing decomposition methods use spectral information to reduce the number of estimated materials, but they often require user-defined thresholds or rely on hard thresholding for segmentation \cite{fredette2019multi, bateman2013segmentation, le2011segmentation, alessio2013quantitative}. Here, we propose a practical approach for spectral distortion correction and inter-pixel variability for any chosen PCD. We extend this empirical method proposed in projection domain by Jakubek et al.~\cite{jakubek2007data} to a computed tomographic setting. This method does not necessitate a-priori knowledge of the detector's response function. This advantage is particularly useful since obtaining such information can be difficult due to pixel-by-pixel variations due to integrated circuits and readout electronics being independent for each pixel in a PCD.

Our proposed Iterative Clustering Material Decomposition (ICMD) method combines an empirical spectral correction method and multi-step material decomposition using cluster analysis, enabling accurate material quantification and classification in a micro CT setting. The spectral distortion correction applied before clustering is critical for achieving quantitative accuracy, given the close similarity in attenuation properties of various materials. 

\begin{figure}[!hb]		
	\centering
	\includegraphics[width=0.5\textwidth]{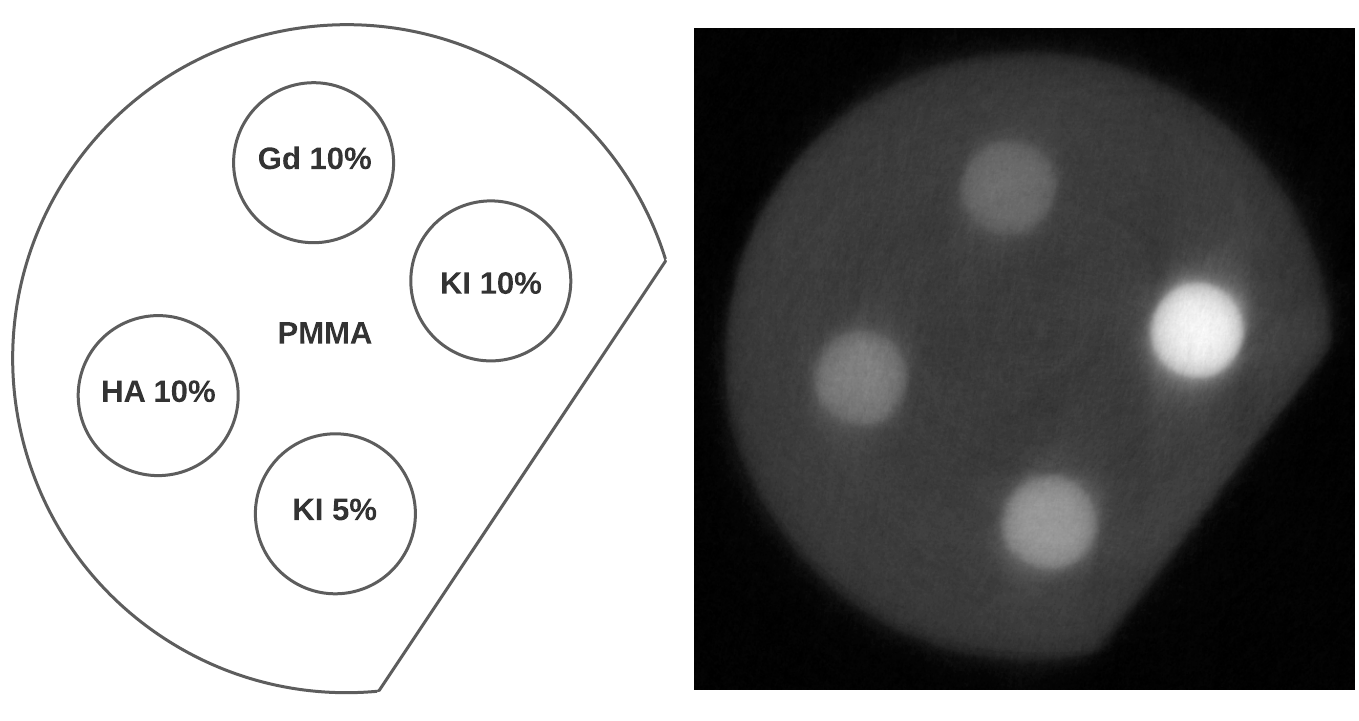}
	\caption{\textcolor{black}{(Left) Cross-sectional schematic of a cylindrical PMMA phantom (27mm diameter) with four cylindrical holes (4mm diameter each). The holes contain 10\% potassium iodide ($\ce{KI}$), 5\% potassium iodide ($\ce{KI}$), 10\% Gadopentetic acid ($\ce{C14H20GdN3O10-H2O}$), and 10\% hydroxyapatite ($\ce{Ca5(PO4)3OH}$) weight percent solution in water, respectively. (Right) A slice of the reconstructed CT slice.}	}
	\label{fig:pmma_phantom}
\end{figure}

\section{Materials and Methods}
In this study, we employ Medipix3 PCD, a key component of the spectral CT system under investigation. This section provides an in-depth description of the Medipix3 PCD, followed by an overview of our micro CT system, including the spectral correction method utilized in tomography to obtain corrected mass attenuation coefficients for each pixel. Finally, we discuss the clustering approach implemented to reduce the number of basis functions used in each iteration of the material decomposition process.

\begin{figure}[!h]
\centering
\includegraphics[width=0.5\textwidth]{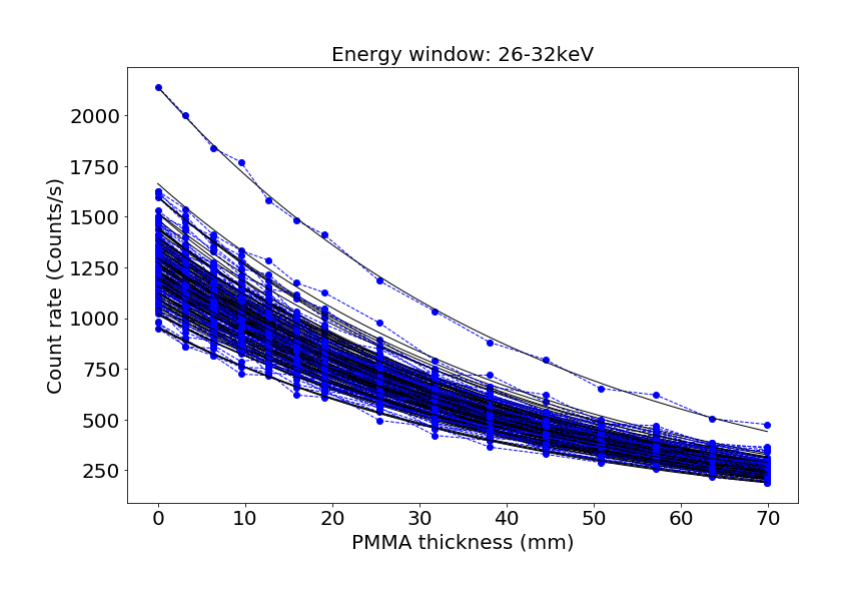}
\caption{Signal-to-thickness calibration for 100 randomly selected pixels based on 15 PMMA thickness measurements, spanning 0.0 to 70.0mm. The calibration curves, set within a 26-32 keV energy window and a 1 second/frame acquisition rate, highlight the variability in inter-pixel response for the same photon counting detector.}
\label{fig:equivalent_PMMA}
\end{figure}

\begin{figure*}[!h]
\centering
\includegraphics[width=0.90\textwidth]{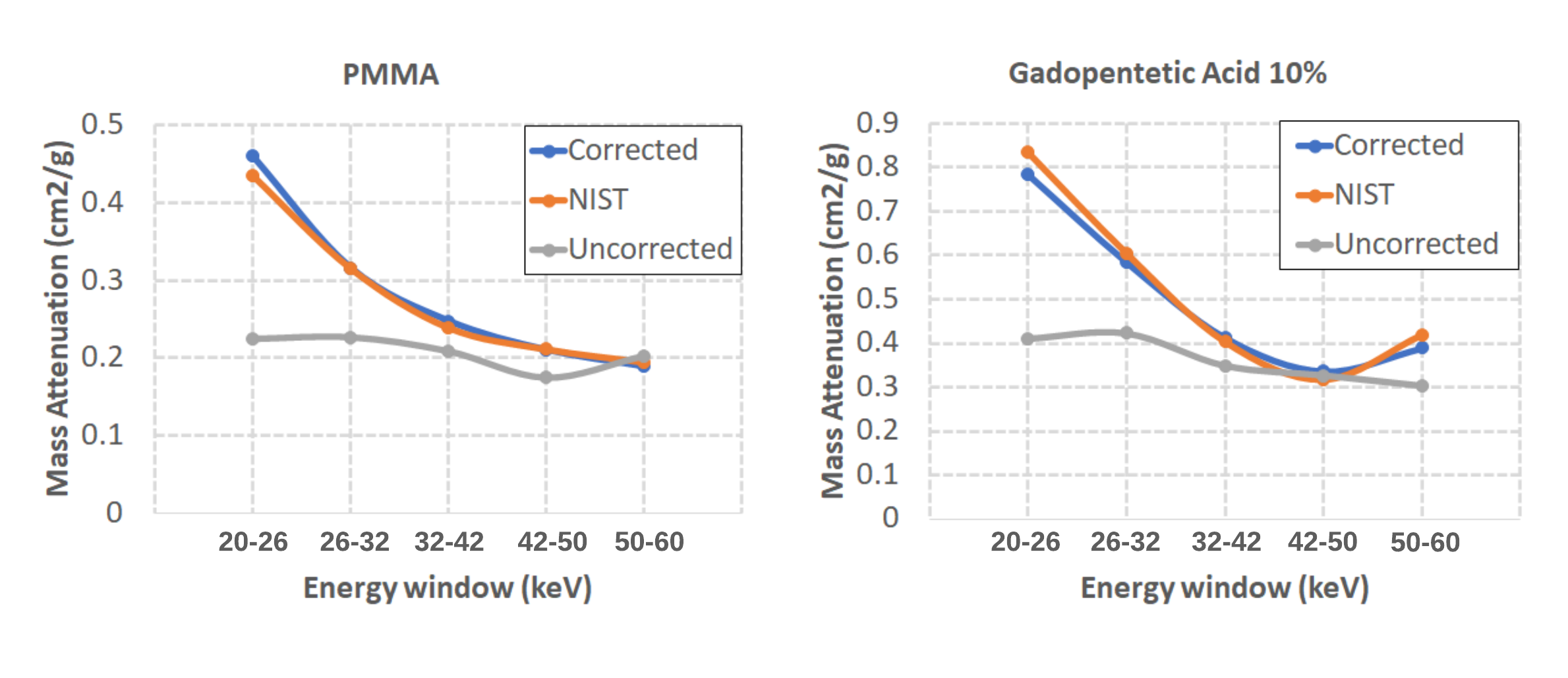}
\caption{This figure displays a comparison of the corrected and uncorrected mass attenuation coefficients from reconstructed voxels against the expected values obtained from the NIST database. The plot demonstrates a strong agreement between the corrected mass attenuation values and the NIST expected values, suggesting a high level of accuracy in the applied correction method.}
\label{fig:mass_attenuation_comparison}
\end{figure*}

\subsection{Medipix3 Photon-counting Detectors (PCDs)}
Conventional x-ray detectors used in clinical settings are energy integrating.  Some avenues for energy discrimination include a dual-energy approach, such as kVp switching, dual source, or source filtering. However, these methods only provide incomplete spectral separation and also often adds to radiation dose due to the need for multiple exposures. 

PCDs are direct conversion detectors that estimate each photon's spatial and spectral information, making them highly attractive for the next generation of biomedical imaging devices. When incoming X-ray photons reaches the semiconductor sensor material, electron-hole pairs (proportional in number to the photon energy) are dislodged from the semiconductor lattice, creating a charge cloud. Under an electric field generated by an externally applied voltage bias, the charged particles drift towards the opposite electrodes. The transient current pulse produced by the displacement of the electron-hole pairs is then processed by the detector's application-specific integrated circuit (ASIC), which includes signal amplification, pulse comparator, and digital counters. The voltage pulse generated is compared with an electronic threshold specified by the user to determine if a count is detected. If the incoming voltage pulse-height exceeds the threshold value, the corresponding digital counter is triggered, and a count is registered. Thus an energy response calibration between pulse height and photon energy is necessary~\cite{ren2018tutorial, vespucci2018robust, 10.1117/12.2082979} and allows distributing the incoming photons into multiple energy bins based on the number of electronic thresholds available for each pixel. A more detailed description of PCDs can be found elsewhere~\cite{ballabriga2018asic, taguchi2013vision}.

The lateral spreading of charge cloud as it drifts through the sensor can result in charge spillover into multiple pixels, reducing spatial and spectral resolution. This effect is known as charge sharing. The new Medipix3 detector features a charge summing mode (CSM), which provides a hardware-level algorithmic correction for charge-sharing~\cite{ren2018tutorial}. Simultaneous counts (from the same incoming photon) in neighboring pixels are collected, and the total charge is allocated to the primary pixel which receives the biggest charge packet . This significantly reduces spectral errors due to charge sharing~\cite{gimenez2011study, ballabriga2018asic}.

The current limitations in semiconductor fabrication technology prevent the growth of large semiconductor crystals with high atomic numbers. In addition, the readout electronics become demanding for a multi-energy detector when constructed as a single large detector with high resolution. Nonetheless, seamless tiling of several smaller detector chips in an array configuration is being investigated to increase the maximum area the device can capture. Our work uses the Medipix3RX line of photon counting detectors developed at CERN (Geneva, Switzerland). This design also enables one of the highest resolution PCDs available to date. These direct conversion semiconducting detectors feature a 1mm thick CdTe sensor with bump bonding at 55$\mu m$ intervals, resulting in virtual pixelation of 55$\mu m$ $\times$ 55$\mu m$. Each individual chip measures 1.408cm $\times$ 1.408cm. The large area Widepix detector (from ADVACAM S. R. O) used in this work is a tiled unit of five individual chips in a row, resulting in an active area of 1.408cm $\times$ 7.040cm, see Fig.~\ref{fig:Widepix}.

\begin{figure}[!h]
\centering
\includegraphics[width=0.5\textwidth]{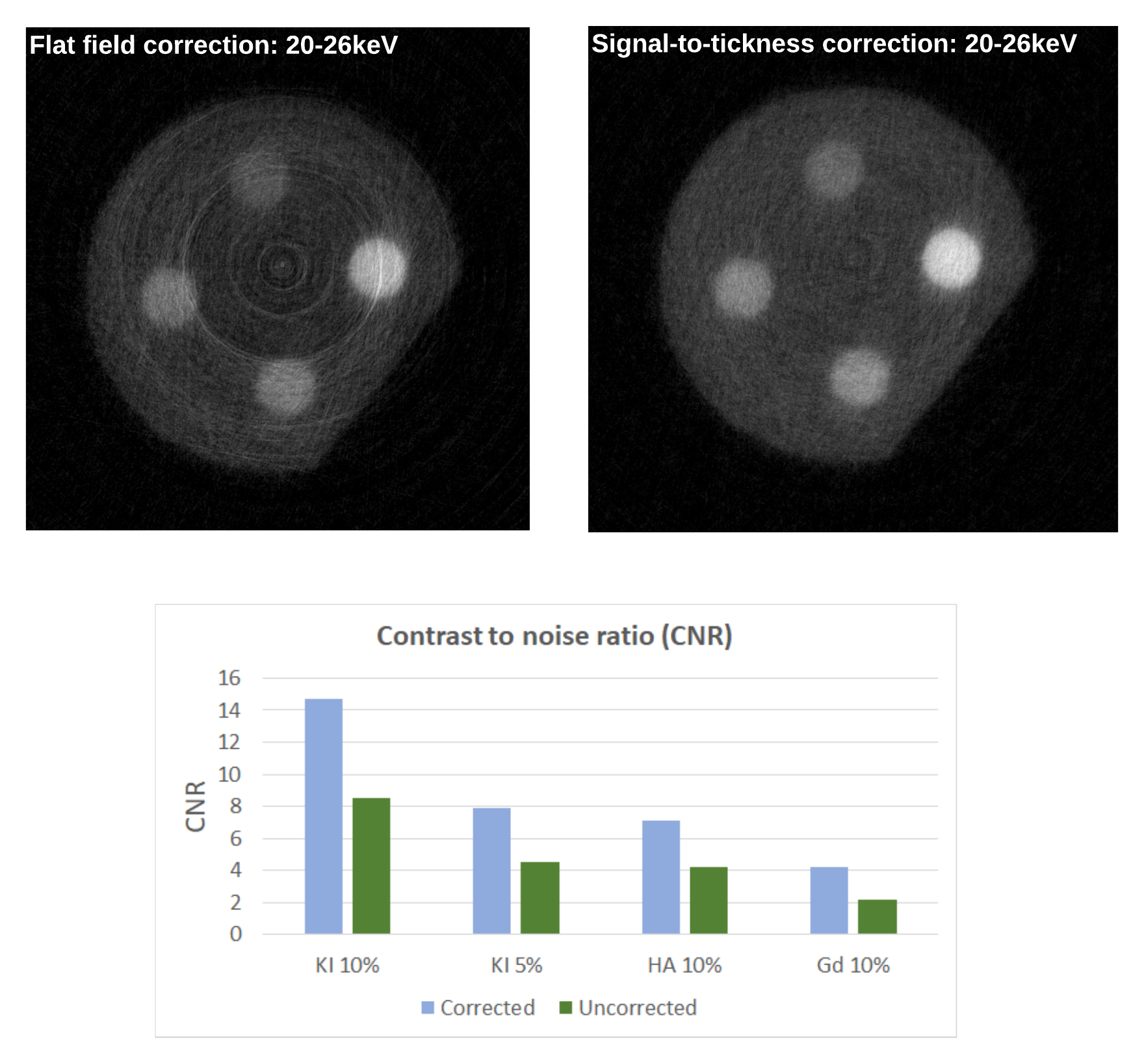}
\caption{Comparison of Contrast to Noise Ratio (CNR) with (right) and without (left) signal-to-thickness calibration. The CNR was calculated as the difference in signal intensity between the region of interest and the PMMA background, normalized by the noise variance in the PMMA background region. Our results show significant improvement in CNR which indicates a better image quality.}
\label{fig:correction_comparison}
\end{figure}

\subsection{Benchtop CT System and Materials}
Our experimental setup comprised of a polychromatic microfocus x-ray tube (Hamamatsu L8121-03) for illumination, with a constant tube potential of 100 \textcolor{black}{kVp}, a current of 500$\mu$A, and a 1.89mm aluminum filter. The x-ray source, operated with a 70$\mu$m focal spot size, was coupled with a Medipix3RX CdTe chip-based wide-area photon-counting detector (tiled array of five individual chips making ~1.4cm X 7cm). Additionally, our CT system incorporated a motorized rotary stage. The source-to-detector distance (SDD) was established at 70cm, while the source-to-object distance (SOD) was set at 60cm. During the CT data acquisition, we maintained the detector in CSM mode with a bias voltage of -540V.

We designed a four-material phantom consisting of a Poly-methyl methacrylate ($\ce{PMMA}$) cylinder filled with vials of 10\% potassium iodide ($\ce{KI}$), 5\% potassium iodide ($\ce{KI}$), 10\% Gadopentetic acid ($\ce{C14H20GdN3O10-H2O}$), and  hydroxyapatite ($\ce{Ca5(PO4)3OH}$) \textcolor{black}{ with 10\% weight solution in water}, as illustrated in Fig~\ref{fig:pmma_phantom}. These materials, relevant to medical imaging, have found utility in numerous material decomposition studies~\cite{fredette2019multi, symons2017photon, le2011segmentation}. Hydroxyapatite forms the main component of bone and calcifications, while iodine and gadolinium serve as common contrast agents in biomedical imaging due to their characteristic K-edge residing within the range of typical medical imaging energies. The PMMA phantom was cylindrical, with a diameter of 27mm and four cylindrical holes, each with a 4mm diameter.

\begin{figure*}[htbp]
  \centering
  \begin{minipage}{0.45\textwidth}
    \includegraphics[width=\textwidth]{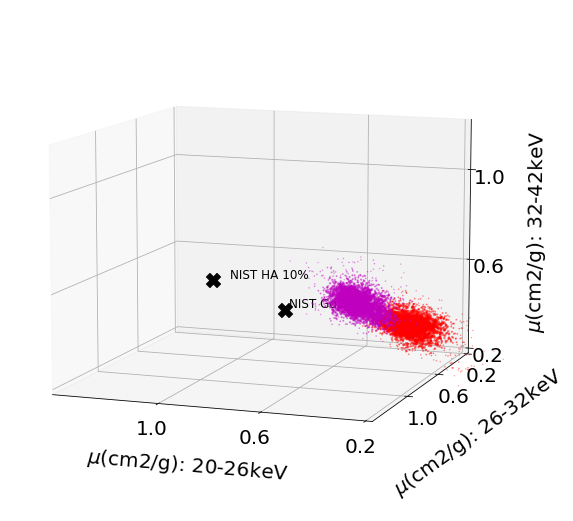}
  \end{minipage}\hfill
  \begin{minipage}{0.45\textwidth}
    \includegraphics[width=\textwidth]{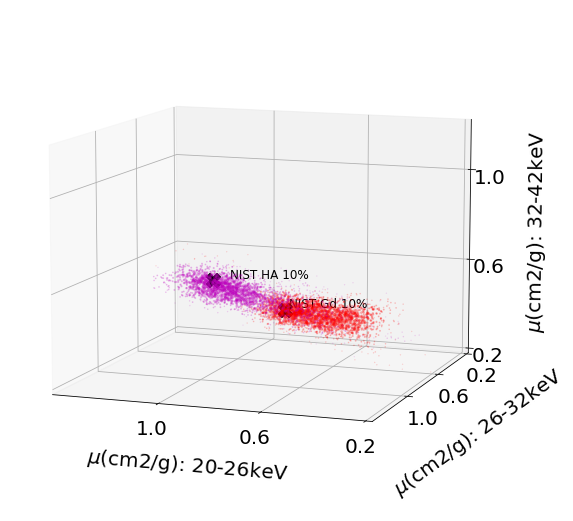}
  \end{minipage}
  \caption{Comparison of uncorrected (left) and corrected (right) mass attenuation values for 10\% gadolinium (red) and hydroxyapatite (magenta) water solutions. Post-spectral correction, the measured values align more closely with the theoretical NIST values (marked with X). }
  \label{fig:combined_image}
\end{figure*}

Employing the energy resolving capabilities of the WidePix (CdTe-1mm) detector, we procured five 3D images, each corresponding to one of the five energy windows utilized in the measurement: 20-26, 26-32, 32-42, 42-50, and 50-60 keV. The energy windows were judiciously selected not only for equal flux but also to capture the full extent of the K-edge effects of iodine and gadolinium. 

\begin{figure}[!ht]
\centering
\includegraphics[width=0.5\textwidth]{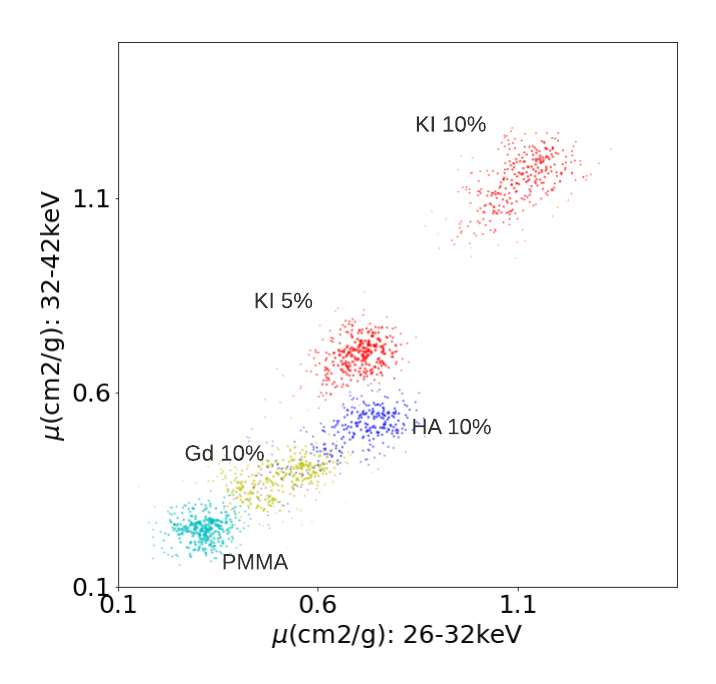}
\caption{A scatter plot of classification (after the implementation of spectral corrections and calibrations) based on  five energy windows.}
\label{fig:all_clusters}
\end{figure}

We collected 230 projection images over a 230-degree span, utilizing a 1-second/frame acquisition time. The total radiation dose across all five energy windows was measured to be 47 mGy using a 10X6-6M mammography ion chamber (Radcal). The resultant CT slice, as depicted in Fig~\ref{fig:pmma_phantom}, represents the mean PMMA thickness when measured across the five energy windows.

\subsection{Empirical Spectral Correction}
\label{sec:spectral_correction}

To address the spectral distortions in PCDs, we implemented an empirical correction method. This involves a signal-to-thickness calibration (STC) and mass attenuation correction described below.

\subsubsection{Obtaining Calibration Parameters}
\textcolor{black}{The initial step in our methodology involved STC. We calibrated the PCD response by assessing the spectral attenuation of uniform PMMA slabs with predetermined thickness values~\cite{jakubek2007data}. PMMA was selected because its attenuation properties closely resemble those of soft tissue. For each of the five energy windows, we took 15 projection measurements at varying PMMA thicknesses, using a 1s acquisition duration per frame, see Fig.~\ref{fig:equivalent_PMMA}. The calibration incorporated PMMA thicknesses ranging from 0 to 70mm, encompassing the entire x-ray attenuation spectrum of our usual samples. Following this, we derived a correlation between PMMA thickness and the observed photon count rate \(I_E^{*}(T)\) via an exponential model:}

\begin{equation} \label{eq:STC1}
I_E^{*}(T) = A_E e^{-B_E T}
\end{equation}

In this equation, $A_E$ and $B_E$ represent the fitting parameters for energy window $E$, with $T$ symbolizing the thickness of the PMMA slab. Note that $A_E$ is dimensionless, whereas $B_E$ has units of reciprocal millimeters .

\begin{equation}
    \ln \left\{ I_E^{*}(T) \right \} = \ln{A_E} - B_E T
\end{equation}

Knowing the detector counts for each pixel for chosen PMMA thickness $T$,  we can obtain the calibration parameters {${A_E, B_E}$} independently for each pixel, using a linear fit.

\subsubsection{Performing STC on Raw CT Data}
\textcolor{black}{Once the calibration parameters were established, we applied STC to the raw projection images from the CT dataset. For each pixel and specific energy window $E$, the observed count rate $I_E$ (corresponding to our sample under investigation) was translated into an equivalent PMMA thickness, denoted as $\delta_E^{PMMA}$. This conversion follows the relationship:}

\begin{equation}\label{eq:STC2}
\delta_E^{PMMA} = \frac{1}{B_E} \ln\left ( \frac{I_E}{A_E} \right )
\end{equation}

\textcolor{black}{Each conversion utilized the respective calibration parameters {${A_E, B_E}$} corresponding to the energy window in question.}

\subsubsection{Reconstruction of Computed Tomography with Equivalent PMMA Thickness}

After applying STC to the raw CT projection data, we perform a 3D reconstruction of the equivalent PMMA thickness (as determined in projection domain from the preceding step). For this reconstruction, we utilized the GPU-accelerated version of the SIRT3D CUDA algorithm, implemented in the \textit{All Scale Tomographic Reconstruction Antwerp} (ASTRA) toolbox, combined with a 3D cone geometry approach~\cite{van2016fast, van2015astra}. The final CT reconstruction provides $\delta_E^{PMMA}$ values for each voxel.
Projections from each energy window $E$ was independently reconstructed, resulting in five separate CT reconstructions corresponding to the five energy windows.

\subsubsection{Mass Attenuation Correction}

In the final step, using the following equivalence, we obtained the corrected mass attenuation values $\mu_E$ for each voxel from the equivalent PMMA thickness $\delta_E^{PMMA}$ obtained from the CT reconstruction 

\begin{equation} \label{eq:attenuation_correction1}
\delta^{\text{VOXEL}} \quad \mu_E = \delta_E ^{PMMA} \quad \mu_E ^{PMMA}
\end{equation}

In this equation, $\delta^{\text{VOXEL}}$ is a constant representing the voxel length (0.055 mm) in the CT reconstruction, $\mu_E$ denotes the mass attenuation ($\textnormal{cm}^2/\textnormal{g}$) of an unknown material, $\delta_E ^{PMMA}$ signifies the equivalent PMMA thickness (cm), and $\mu_E ^{PMMA}$ represents the theoretical mass attenuation ($\textnormal{cm}^2/\textnormal{g}$) of PMMA, as sourced from the National Institute of Standards and Technology (NIST).

Utilizing Eqs.~\ref{eq:STC2} and \ref{eq:attenuation_correction1}, the corrected linear mass attenuation coefficients of an unknown material for energy window $E$ can be computed for each voxel as:

\begin{equation} \label{eq:attenuation_correction2}
\mu_E = \frac{\frac{1}{B_E} \ln\left ( \frac{I_E}{A_E} \right) }{\delta^{\text{VOXEL}}} \quad \mu_E^{PMMA}
\end{equation}

\textcolor{black}{This correction is applied independently to each energy window $E$ utilizing its respective calibration parameters {${A_E, B_E}$}.}

\begin{figure*}[!ht]
\centering
\includegraphics[width=0.9\textwidth]{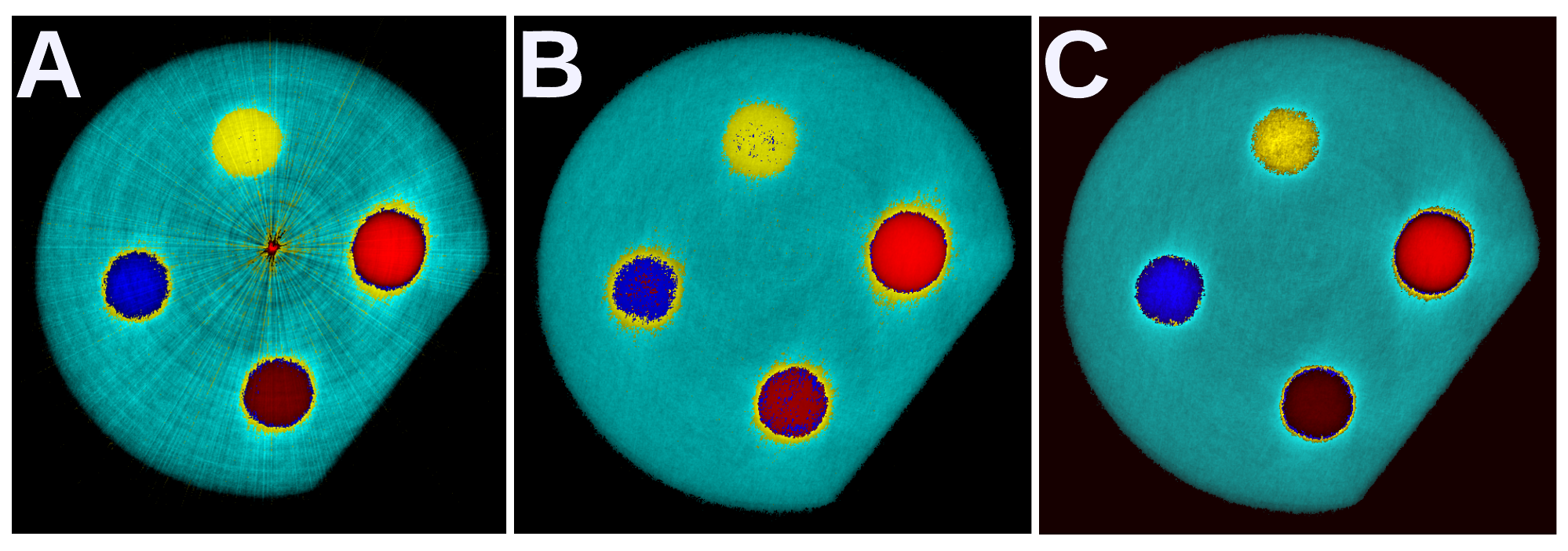}
\caption{Illustration of fundamental clusters derived under three different conditions. A) Flat-field correction applied to projection images with voxel clustering across all five energy windows. B) With our empirical correction method but only using two energy windows: 26-32 and 32-42 keV. C) With our empirical correction method using all five energy windows. Clusters are represented as: PMMA (cyan), potassium iodide (red), Gadopentetic acid (yellow), and Hydroxyapatite (blue).}
\label{fig:clusters}
\end{figure*}

\begin{figure}[!h]
\centering
\includegraphics[width=0.7\textwidth]{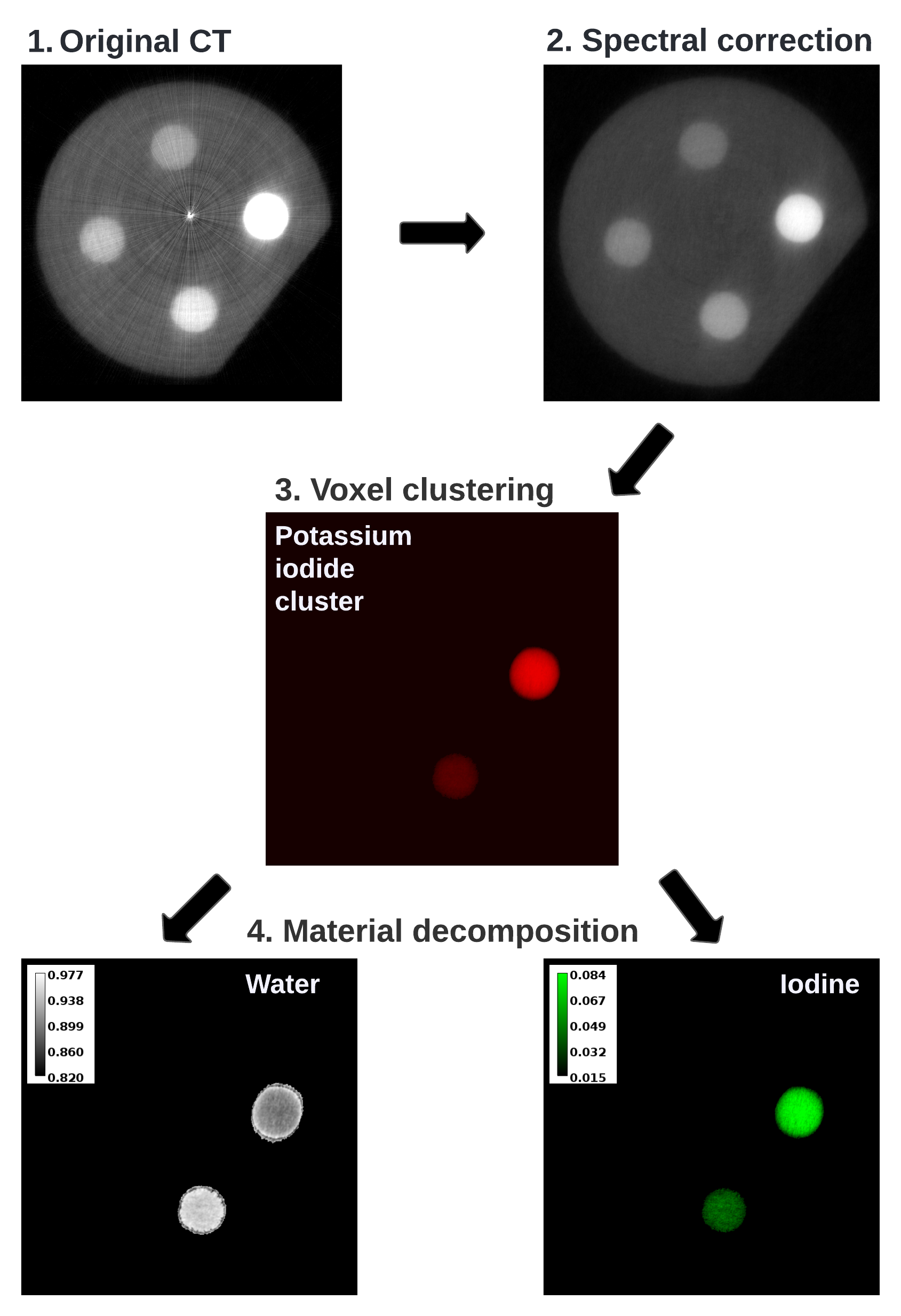}
\caption{This diagram illustrates the data processing pipeline. The first step is the correction of spectral distortions that is effective in the reconstruction domain. The second step involves iterative cluster analysis where each cluster has fewer unknowns to be solved for. In the final step, decomposition methods are applied to each cluster at a time (with reduced basis materials), and the mass fraction of each material within each cluster is determined using least squares. In steps 3-5, the decomposition of the potassium iodide cluster into its water and iodide components is highlighted as an example. Steps 3-5 are applied to each individual cluster obtained after step2. }
\label{fig:data_processing_pipeline}
\end{figure}

\subsection{Voxel Clustering Using Spectral Information}
Before applying material decomposition methods, we use clustering analysis to separate voxels to a few possible classes or clusters. The decomposition is then applied to each cluster with reduced basis functions or unknowns. In this paper, we define clustering as the process of assigning individual CT voxels to different groups based on their measured mass attenuation values across various energy windows. For each voxel, there is a vector $\vec{x} = (\mu_1,...,\mu_m)$, where $\mu_m$ represents the corrected mass attenuation coefficient for the $m^{th}$ energy window. However, the measured attenuation properties of biological tissue are too heterogeneous and dispersed to be accurately captured by threshold-based segmentation techniques. We examine the use of an unsupervised machine learning method, such as Gaussian mixture models (GMM), where the model learns from the intrinsic structure of the data without requiring training~\cite{reynolds2009gaussian}.

GMM assumes that the data is generated from a mixture distribution with $K$ components (number of tissue types). For a Gaussian mixture $p(\vec{x})$ with $K$ components, the $i^{th}$ component has a mean of $\widetilde{\mu}_i$, weight $\pi_i$, and a covariance matrix $\Sigma_i$. The probability $p(\vec{x})$ for each element $\vec{x} = (\mu_1,...,\mu_m)$ is given by

\begin{equation} \label{eq:mixture1}
p(\vec{x}) = \sum_{i=1}^K \pi_i f_i (\vec{x} | \widetilde{\mu_{i}}, \Sigma_i)
\end{equation}

\begin{equation} \label{eq:mixture2}
f(\vec{x} | \widetilde{\mu_{i}}, \Sigma_i) = \frac{1}{\sqrt{(2 \pi)^K |\Sigma_i|}} \textbf{exp} \left( -\frac{1}{2} (\vec{x} - \widetilde{\mu_{i}})^T \Sigma_{i}^{-1} (\vec{x} - \widetilde{\mu_{i}}) \right)
\end{equation}

\begin{equation} \label{eq:mixture3}
\sum_{i=1}^{K} \pi_i = 1
\end{equation}

When the number of components $K$ is known, the mixture model parameters ${ \pi_i, \widetilde{\mu_{i}}, \Sigma_i }$ can be estimated using the expectation maximization technique \cite{pan2002maximum}. Using these estimated model parameters, we can now calculate the probability that a data point $\vec{x}$ belongs to cluster $C_i$:

\begin{equation} \label{eq:probability}
p(C_i | \vec{x}) = \frac{\pi_i f_i (\vec{x} | \widetilde{\mu_{i}}, \Sigma_i)}{\sum_{i=1}^K \pi_i f_i (\vec{x} | \widetilde{\mu_{i}}, \Sigma_i)}
\end{equation}

\textcolor{black}{We employed the \texttt{scikit-learn} library\cite{scikit-learn} in Python to perform GMM classification. Our implementation included full covariance, a specified random seed, and a $0.01$ tolerance threshold for the stopping criteria} \\ 

We start by employing the GMM model to classify the entire CT voxel dataset based on the corrected mass attenuation values for all energy bins. Each clustering stage separates the chosen voxel sets to two clusters. The most identifiable cluster (based on proximity to NIST value for all energy bins available) is separated out and the clustering process is repeated on the second cluster. This iterative clustering is repeated until until the identification of "fundamental" clusters corresponding to each class of material, beyond which no further distinctions can be made. 

Once every voxel in the CT dataset is clustered, the next step involves recognizing and labeling these clusters. This is achieved by comparing the average mass attenuation value of each cluster against theoretical NIST values of the unknown materials. 

\subsection{Material Decomposition }
\label{sec:material_decomposition}

The mass attenuation characteristic of each voxel can be modeled as a linear combination of the materials constituting the volume~\cite{le2011least, le2011segmentation, lee2014quantitative}. The mass attenuation, \(\mu_M\), of such a mixture is given by:

\begin{equation} 
\mu_M(E) = \sum_{i=1}^N m_i  \mu_{i}(E)
\label{eq:material_decomposition_1}
\end{equation}

In Eq.~\ref{eq:material_decomposition_1}, \(m_i\) denotes the mass fraction of each material in the mixture and \(\mu_i\) denotes the mass attenuation of the material basis which is obtained based on the known values from the NIST database. Concurrently, the summation of the mass fractions for each material equates to one:

\begin{equation} 
\sum_{i=1}^N m_i = 1, \quad i =1, ..., N
\label{eq:material_decomposition_2}
\end{equation}

This constraint presented in Eq.~\ref{eq:material_decomposition_2} ensures the conservation of mass across the mixture.

In a single-step material decomposition method, one attempts to solve for all of the unknowns in one step using Eqs.~\ref{eq:material_decomposition_1} and \ref{eq:material_decomposition_2} above, when reconstructed image volumes corresponding to multiple energy bins are available using PCDs or other spectral imaging methods. Thus for each voxel in the CT reconstruction, one assumes that all of the \(N\) materials could be possible unknowns, making the problem usually very ill-conditioned. 

In our ICMD, we first separate the voxels into various clusters using GMM after careful corrections for mass attenuation values in each CT voxel.  Following this, the decomposition step is applied to each cluster in an iterative fashion where we now have limited basis functions making the problem less ill-conditioned. An example with results is shown in the sections below.

\section{Results and Discussion}

\subsection{Signal-to-Thickness Calibration and Mass Attenuation Correction}
The initial step in our data processing pipeline involves acquiring multi-energy computed tomography measurements, resulting in a set of projections for each energy window. By employing a signal-to-thickness calibration method, these projections are subsequently converted to equivalent PMMA thickness values using Eq.~\ref{eq:STC2}. The resulting PMMA equivalent projections are then reconstructed to generate 3D multi-energy images of the object under investigation.\\
After STC, we performed mass attenuation correction. We utilize Eq.~\ref{eq:attenuation_correction2} to convert the equivalent PMMA thickness of each voxel into mass attenuation values. To verify the efficacy of our correction approach, we juxtaposed the measured mass attenuation values of two known materials with their respective NIST values, as depicted in Fig.~\ref{fig:mass_attenuation_comparison}. The corrected mass attenuation was found to align accurately with the expected theoretical NIST values.  \\

\begin{table}[!h]
\centering
\includegraphics[width=0.8\textwidth]{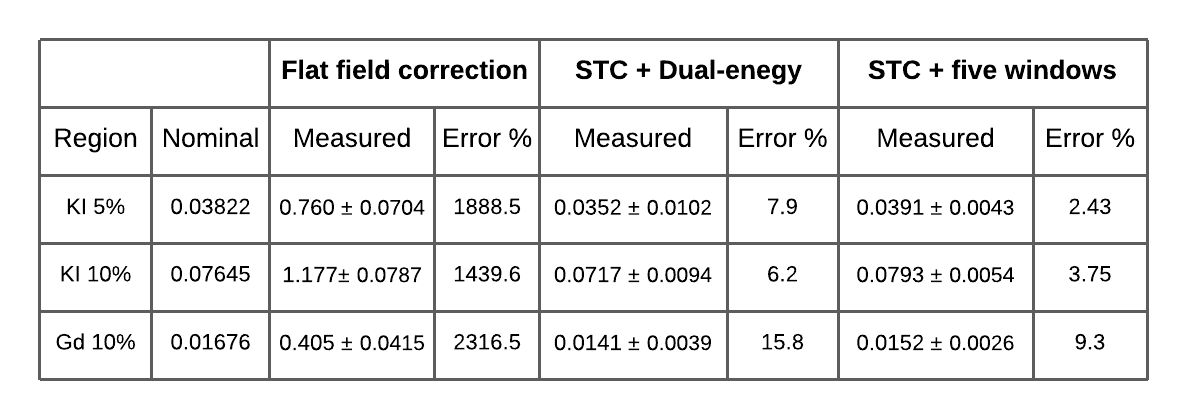}
\caption{Dual-contrast agent mass-fraction values for iodine(I) and gadolinium(Gd).}
\label{tbl:mass_fractions}
\end{table}

Our correction method also significantly impacts the noise properties of the reconstructed CT images. CT reconstruction artifacts, such as ring artifacts, often emerge from variations in pixel sensitivity across different energy windows. These variations primarily stem from inter-pixel variability and also due to physical effects such as beam hardening. Attenuation inhomogeneities across the sample lead to variations in the spectra that reach each detector pixel. Given that the detection efficiency is energy-dependent, this results in each pixel having a slightly varying overall absorption efficiency based on the local attenuation properties of the sample. Traditional flat-field correction methods, which assume a uniform energy response across all pixels, are insufficient for correcting these variations.\\
Our STC has shown to be highly effective in enhancing the signal-to-noise ratio (SNR), with improvements close to doubling the original value, as demonstrated in Fig.~\ref{fig:correction_comparison}. The SNR was calculated as  $SNR = \frac{\mu_{\text{signal}} - \mu_{\text{PMMA}}}{\sigma_{\text{PMMA}}}$, where $\mu_{\text{signal}}$ represents the mean pixel intensity of the region of interest (the signal), $\mu_{\text{PMMA}}$ is the mean pixel intensity of the background (PMMA), and $\sigma_{\text{PMMA}}$ is the standard deviation of the pixel intensities in the PMMA background (the noise).

\begin{figure}[!h]
\centering
\includegraphics[width=0.5\textwidth]{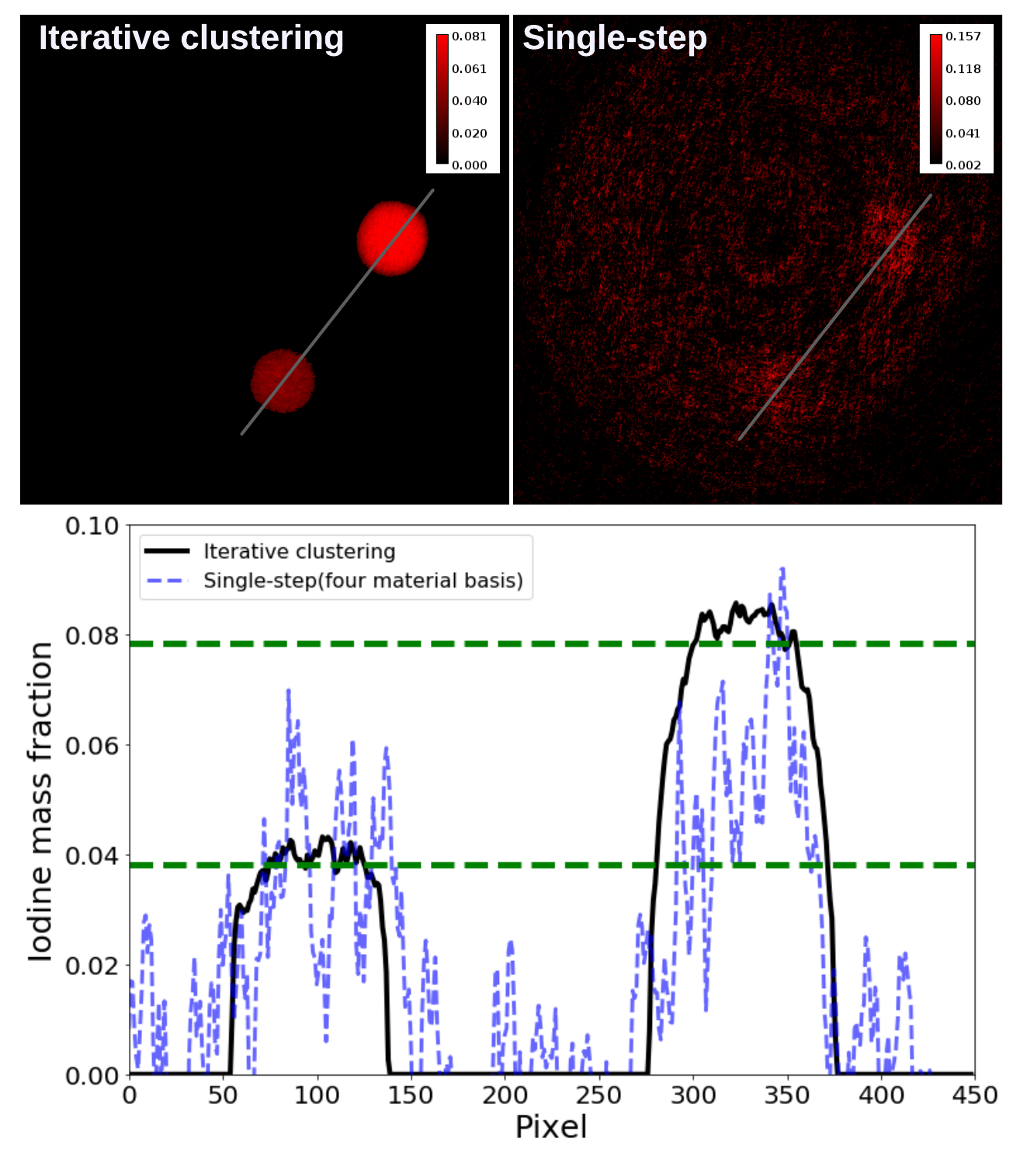}
\caption{Contrasting our ICMD method with single-step material decomposition using a four-material basis simultaneously. This graph underscores the advantages of limiting the number of material bases, which contributes to noise reduction and accuracy enhancement. It provides a clear comparison between our ICMD method and the conventional single-step material decomposition approach.}
\label{fig:line_profile_iodine}
\end{figure}

\subsection{Voxel Clustering Using Spectral Information}
After implementing the signal-to-thickness calibration, we proceed with an iterative voxel clustering with GMM as described in the sections above. In Fig.~\ref{fig:combined_image}, we demonstrate the advantages of our empirical correction method by examining two clusters: gadolinium and hydroxyapatite, within the volume. The left side of the image presents the clustering results achieved with only the flat-field correction. The right side displays the clustering distribution derived from our ICMD method. This comparison underscores the impact of our correction technique on both cluster distributions and their centroids. It is worth noting that while we utilized all five energy bins for the clustering of the data displayed, we have chosen to present only the cluster map associated with three energy bins to streamline visualization. A notable enhancement brought about by our correction method is evident in the precision of the attenuation values. Although there's a range of values associated with each material, the clusters align more closely with the true values after applying our empirical correction technique.\\
For our sample of interest with PMMA background and three distinct signal types and also including different concentrations of one of the signals, the GMM clustering yielded five fundamental clusters corresponding to PMMA, potassium iodide (two distinct concentrations), gadopentetic acid, and hydroxyapatite as identified by their proximity to the respective true mass attenuation values obtained from NIST, see Fig.\ref{fig:all_clusters}. Material decomposition method described in section~\ref{sec:material_decomposition} would then be applied to each of these fundamental clusters to yield quantitatively accurate mass fractions of each material (results to be presented in the next section).

\begin{figure*}[!ht]
\centering
\includegraphics[width=0.8\textwidth]{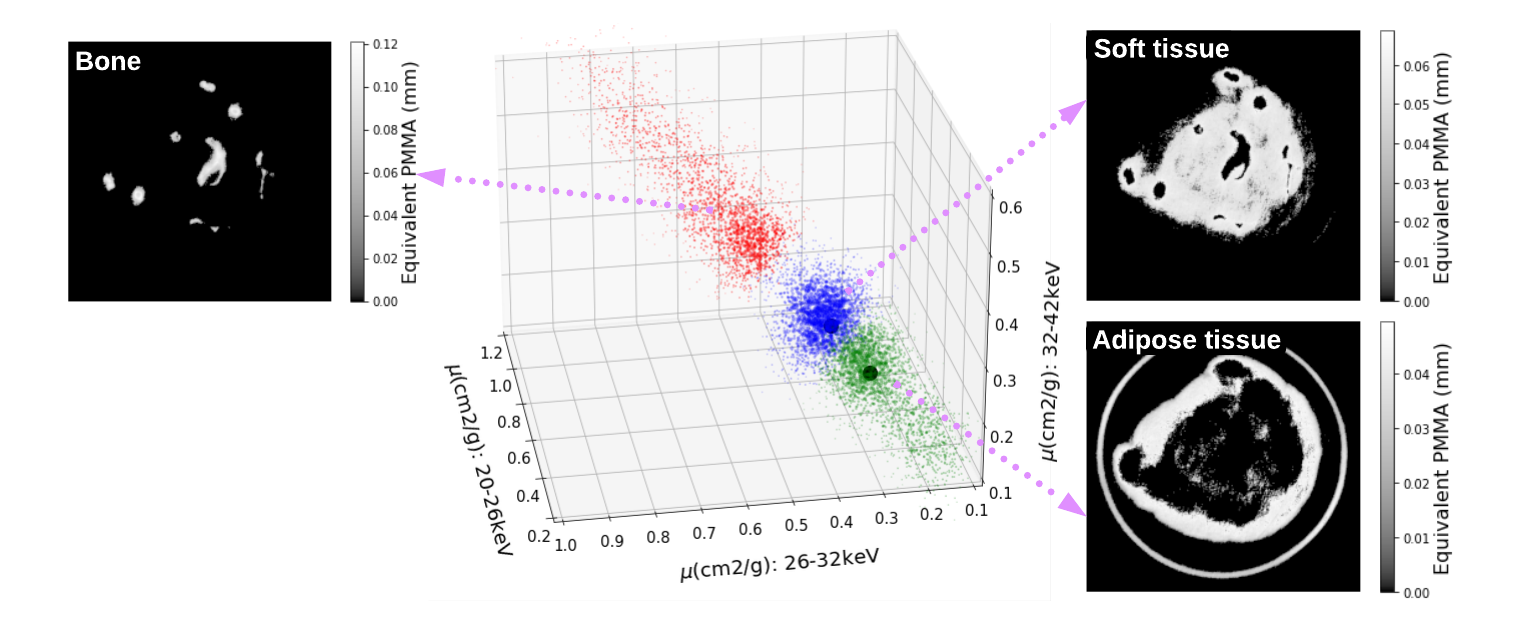}
\caption{Unsupervised segmentation of heterogeneous biological tissues into bone, soft, and adipose types using the Gaussian Mixture Model (GMM). Fundamental clusters were obtained through random initialization, leveraging the distinct mass attenuation values of the tissues.}
\label{fig:mice_tissue_clusters}
\end{figure*}

\begin{figure}[!h]
\centering
\includegraphics[width=0.6\textwidth]{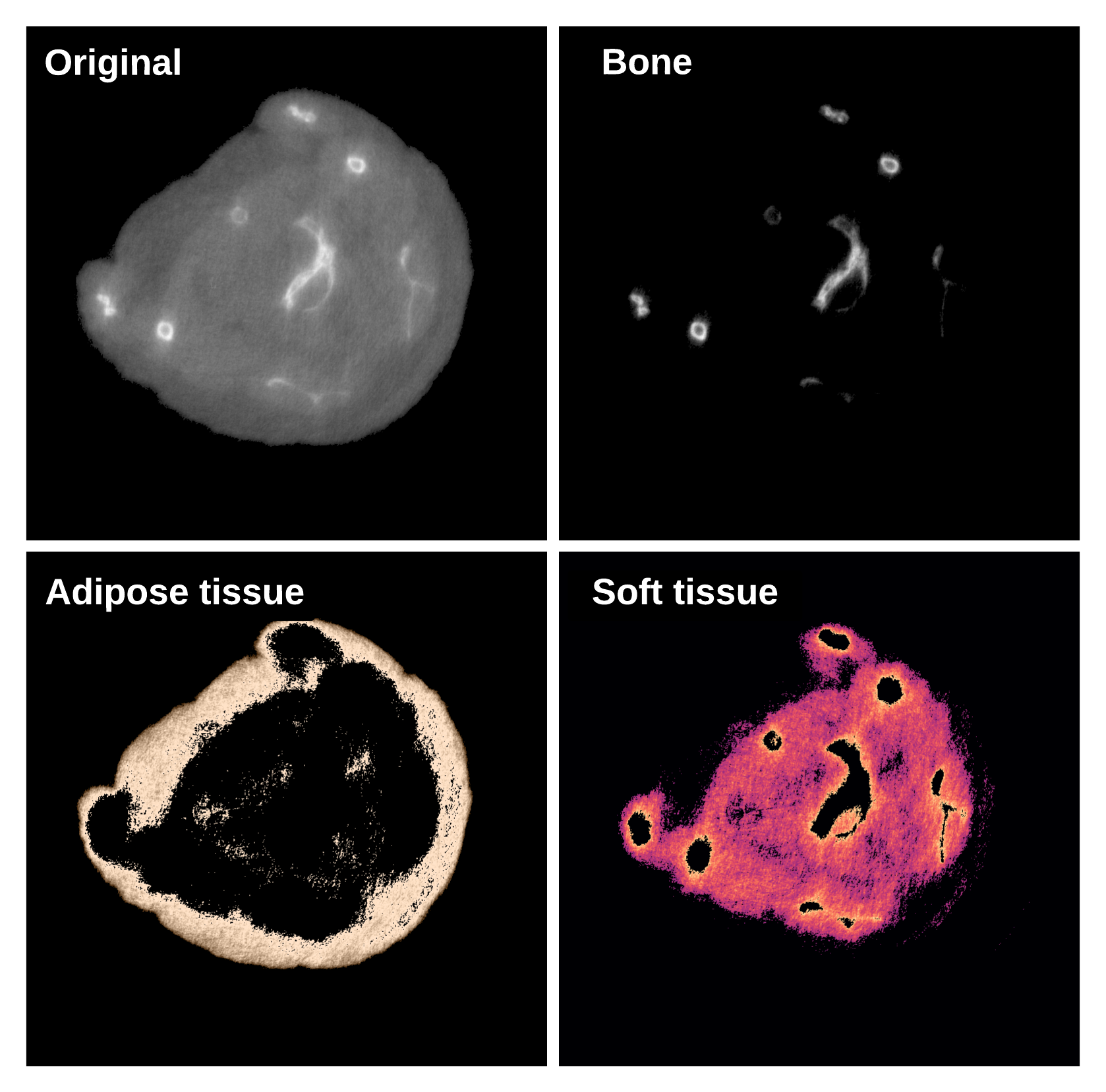}
\caption{Mouse tissue sample processed with our ICMD method. The Gaussian Mixture Model (GMM) identifies three fundamental clusters, corresponding to bone, soft, and adipose tissue.}
\label{fig:tissue_classification1}
\end{figure} 

In Fig.~\ref{fig:clusters}, we show the benefits of our empirical correction method using CT slice with all identified clusters.  
Fig.~\ref{fig:clusters}(\textbf{A}), shows results of clustering when only a flat-field correction was applied to the projection images. One can see that even when we utilized all five energy windows for voxel classification, in this scenario, flat-field correction alone is unable to eliminate CT ring artifacts or ensure accurate classification. In addition, the per-pixel variability heightens the probability of voxel misclassification, leading to noise and incorrectly classified voxels appearing as rings. In Fig.~\ref{fig:clusters}(\textbf{B}), we employed signal-to-thickness calibration in projection but used only two (26-32 and 32-42 keV) of the five available energy windows  for voxel classification. While signal-to-thickness calibration diminished noise artifacts caused by inter-pixel variations, it still led to a considerable number of misclassified voxels. When using signal-to-thickness calibration and voxel classification with all five energy windows, we observed improved voxel clustering  (Fig.~\ref{fig:clusters}(\textbf{C})). However, we still encountered incorrectly classified voxels at the boundary between materials, suggesting that such classification with mass attenuation values alone are not enough for complete tissue classification in a complex mixture of materials.\\

\subsection{Material Decomposition}

Following the clustering step, we employed a least-squares method to solve Eqs.~\ref{eq:material_decomposition_1} and \ref{eq:material_decomposition_2} for a reduced set of possible material bases in each of the cluster. In Fig.~\ref{fig:data_processing_pipeline}, we present a general overview of the data processing pipeline. The material bases for each fundamental cluster were identified using the expected theoretical values from NIST, which allowed us to limit the number of possible material bases to a maximum of three. For material decomposition, we utilized only two energy windows (20-26 and 26-32 keV), deliberately avoiding the iodine K-edge energy range (32-42 keV) as our spectral correction does not fully capture the K-edge effects. Material decomposition steps were applied to each voxel cluster independently to yield quantitatively accurate mass fractions.
 The mass fractions thus obtained for some relevant components in our cylindrical phantom are summarized in Table~\ref{tbl:mass_fractions}. The results also show the advantages of STC and multi-energy windows on clustering over the basic flat-field correction in yielding quantitative accuracy in material decomposition. The results obtained when using only the simple flat-field correction is also presented for comparison. As noted, upon applying our empirical correction, the measured mass fractions closely matched the nominal mass fractions for each one of the solutions, with a maximum error of 3.75\% for iodine and 9.3\% for Gd. We do not present results for the background material PMMA as this material can be readily identified by directly comparing the measured mass attenuation values with the NIST values. Likewise, we did not include results for the hydroxyapatite solution due to the tendency of the solute to sediment at the bottom of the vial, which complicates comparisons with NIST values.

To better visualize benefits in quantitative accuracy and noise reduction yielded with our ICDM method supported by our empirical spectral correction, we show the decomposed mass fractions of iodine obtained using our ICMD using five energy windows for classification in comparison with the single-step material decomposition technique after STC and mass attenuation correction.  As shown in Fig.~\ref{fig:line_profile_iodine}, our method demonstrated significant noise reduction with higher accuracy for both the high and low concentration solutions of iodine.

\subsection{Ex Vivo Tissue Classification}
We applied our techniques to spectral micro-CT data of a mouse, acquired under identical experimental conditions (as described for the PMMA phantom) in our laboratory. The tissue classification results are shown for a slice from the mouse's upper chest region. For consistency, we maintained the same calibration parameters and energy windows as the prior experiment.

Following the signal-to-thickness calibration, utilizing PMMA slabs as outlined in Section~\ref{sec:spectral_correction}, the 3D image of the mouse was processed, leading to an artifact-free reconstruction ready for cluster analysis.

\begin{figure}[!h]
\centering
\includegraphics[width=0.5\textwidth]{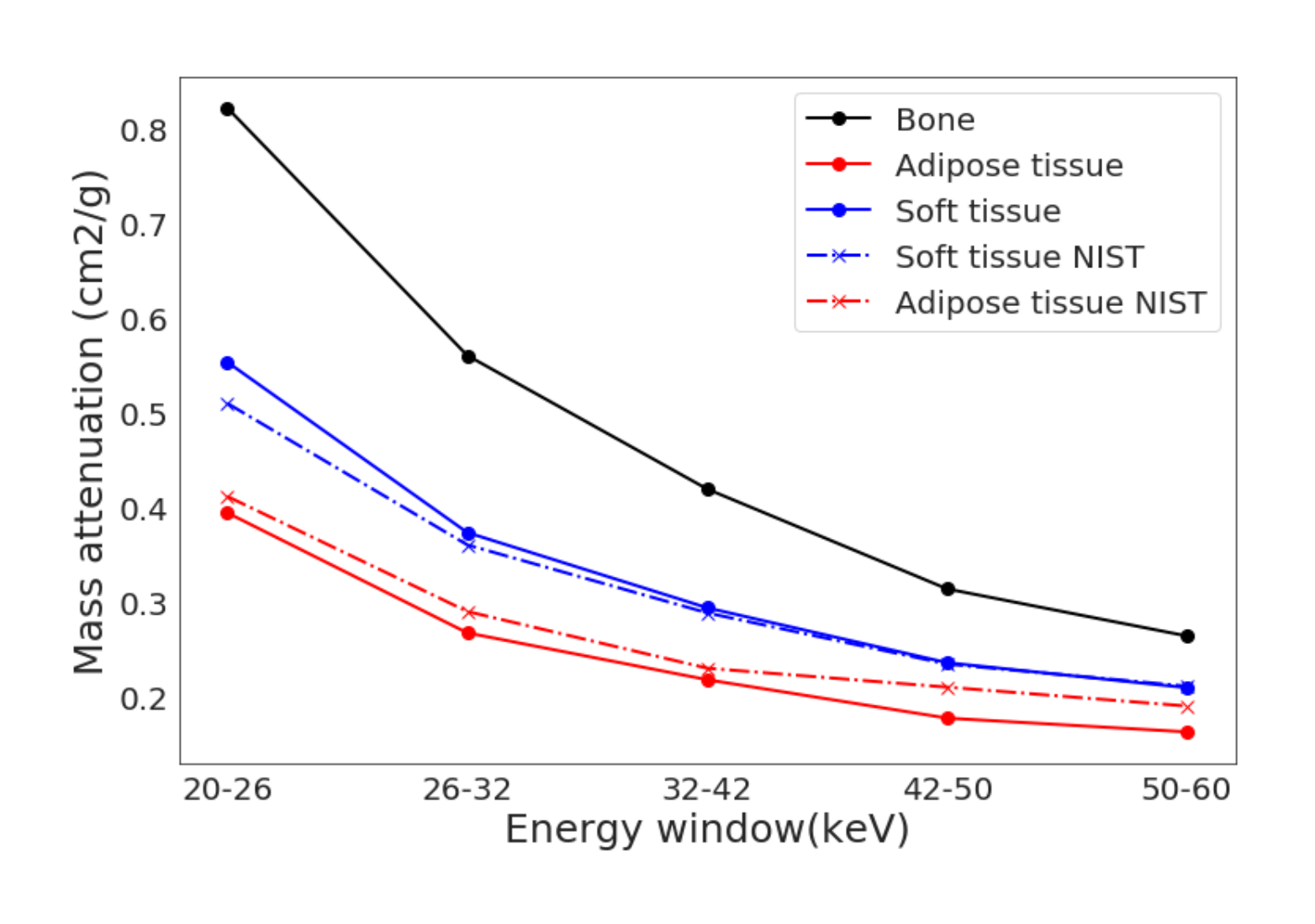}
\caption{Graph depicting measured mass attenuation for soft, adipose, and bone tissues. Experimental points represent mean values of mass attenuation for each cluster at each energy window, while the dotted line indicates expected theoretical values from NIST. Deviations from theoretical values are attributed to the inherent inhomogeneity and complex composition of mouse tissue.}
\label{fig:tissue_classification2}
\end{figure}

Fig.~\ref{fig:mice_tissue_clusters} shows the scatter density plot of the three clusters identified using GMM. In Fig.~\ref{fig:tissue_classification1}, we present the image representation for each of these fundamental clusters. These clusters are representative of bone, soft, and adipose tissue. We show the comparison of the mean mass attenuation values of each cluster with the NIST model for soft and adipose tissues in Fig.~\ref{fig:tissue_classification2}. Due to lack of sufficient and standardized data for spectral mass attenuation values of mouse bone, we have not included standard values for bone in this plot.

\section{Conclusion}

We present an iterative clustering method followed by material decomposition for spectral computed tomography when using a high resolution photon counting spectral detector in a micro CT setup. The correction method requires acquisition of spectral projection data for multiple thicknesses of known materials with the same imaging setting. We describe and demonstrate how this correction applied to individual projections can be translated effectively to the reconstructed CT volume. Our correction method accounts for multiple spectral errors including fluorescence from the sensor material (CdTe) and beam hardening. It also accounts for pixel-to-pixel variations in detector gain. The result is artifact and noise reduction in the reconstructed image along with accurately reconstructed mass attenuation values per voxel and for all energy bins of interest. A major advantage of this correction method is that it does not require a priori knowledge of detector architecture or detector response function, thus making it easy to adapt to any chosen detector.

We show that iterative Gaussian mixture clustering method prior to material decomposition can reduce the number of unknowns in each step of the decomposition applied to one class or cluster at a time. Thus one encounters a better-conditioned material decomposition problem for each cluster in comparison to single-step method where all unknowns are solved in one step. We show the benefit of spectral correction for this classification. For the materials and concentrations we used in our phantom, using five energy bins yielded better classification results than by using only two energy bins. With our ICMD method that involves iterative clustering and material decomposition, we show quantitative accuracy and noise reduction in decomposition for multiple materials and varying concentrations in a micro CT phantom. We also show results from mouse imaging with tissue and bone classification.

 For this work, the spectral data was obtained from the Medipix3 photon counting detector with high resolution (55 microns). Our detector unit was a tiled array of five individual chips making about 1.4cm X 7cm. The tiling process has not contributed to any noticeable artifacts in the microCT images. We will be testing a larger area detector with tiling in 2D for our next prototype unit. 

In summary, we show that by combining empirical spectral correction methods for PCDs and iterative clustering and decomposition methods, one can improve quality and accuracy in multi-material decomposition in spectral CT. While our method is shown with a micro CT setup, these approaches would translate to all spectral CT applications. Future work will examine benefits of improved energy bin selection as well as strategies for dose reduction in spectral micro CT by combining statistical image reconstruction methods\cite{5610726} for our newly proposed ICDM approach. 

\bibliographystyle{ieeetr}
\bibliography{bibliography}

\begin{thebibliography}{10}

\bibitem{tao2019feasibility}
S.~Tao, K.~Rajendran, C.~H. McCollough, and S.~Leng, ``Feasibility of
  multi-contrast imaging on dual-source photon counting detector (pcd) ct: An
  initial phantom study,'' {\em Medical physics}, vol.~46, no.~9,
  pp.~4105--4115, 2019.

\bibitem{symons2017photon}
R.~Symons, B.~Krauss, P.~Sahbaee, T.~E. Cork, M.~N. Lakshmanan, D.~A. Bluemke,
  and A.~Pourmorteza, ``Photon-counting ct for simultaneous imaging of multiple
  contrast agents in the abdomen: an in vivo study,'' {\em Medical physics},
  vol.~44, no.~10, pp.~5120--5127, 2017.

\bibitem{fredette2019multi}
N.~R. Fredette, A.~Kavuri, and M.~Das, ``Multi-step material decomposition for
  spectral computed tomography,'' {\em Physics in Medicine \& Biology},
  vol.~64, no.~14, p.~145001, 2019.

\bibitem{ren2018tutorial}
L.~Ren, B.~Zheng, and H.~Liu, ``Tutorial on x-ray photon counting detector
  characterization,'' {\em Journal of X-ray science and technology}, vol.~26,
  no.~1, pp.~1--28, 2018.

\bibitem{taguchi2013vision}
K.~Taguchi and J.~S. Iwanczyk, ``Vision 20/20: Single photon counting x-ray
  detectors in medical imaging,'' {\em Medical physics}, vol.~40, no.~10,
  p.~100901, 2013.

\bibitem{jakubek2007data}
J.~Jakubek, ``Data processing and image reconstruction methods for pixel
  detectors,'' {\em Nuclear Instruments and Methods in Physics Research Section
  A: Accelerators, Spectrometers, Detectors and Associated Equipment},
  vol.~576, no.~1, pp.~223--234, 2007.

\bibitem{touch2016neural}
M.~Touch, D.~P. Clark, W.~Barber, and C.~T. Badea, ``A neural network-based
  method for spectral distortion correction in photon counting x-ray ct,'' {\em
  Physics in Medicine \& Biology}, vol.~61, no.~16, p.~6132, 2016.

\bibitem{bateman2013segmentation}
C.~J. Bateman, J.~McMahon, A.~Malpas, N.~De~Ruiter, S.~Bell, A.~P. Butler,
  P.~H. Butler, and P.~F. Renaud, ``Segmentation enhances material analysis in
  multi-energy ct: A simulation study,'' in {\em 2013 28th International
  Conference on Image and Vision Computing New Zealand (IVCNZ 2013)},
  pp.~190--195, IEEE, 2013.

\bibitem{le2011segmentation}
H.~Q. Le and S.~Molloi, ``Segmentation and quantification of materials with
  energy discriminating computed tomography: A phantom study,'' {\em Medical
  physics}, vol.~38, no.~1, pp.~228--237, 2011.

\bibitem{alessio2013quantitative}
A.~M. Alessio and L.~R. MacDonald, ``Quantitative material characterization
  from multi-energy photon counting ct,'' {\em Medical physics}, vol.~40,
  no.~3, p.~031108, 2013.

\bibitem{vespucci2018robust}
S.~Vespucci, C.~S. Park, R.~Torrico, and M.~Das, ``Robust energy calibration
  technique for photon counting spectral detectors,'' {\em IEEE transactions on
  medical imaging}, vol.~38, no.~4, pp.~968--978, 2018.

\bibitem{10.1117/12.2082979}
M.~Das, B.~Kandel, C.~S. Park, and Z.~Liang, ``{Energy calibration of photon
  counting detectors using x-ray tube potential as a reference for material
  decomposition applications},'' in {\em Medical Imaging 2015: Physics of
  Medical Imaging} (C.~Hoeschen and D.~Kontos, eds.), vol.~9412, p.~941214,
  International Society for Optics and Photonics, SPIE, 2015.

\bibitem{ballabriga2018asic}
R.~Ballabriga, M.~Campbell, and X.~Llopart, ``Asic developments for radiation
  imaging applications: The medipix and timepix family,'' {\em Nuclear
  Instruments and Methods in Physics Research Section A: Accelerators,
  Spectrometers, Detectors and Associated Equipment}, vol.~878, pp.~10--23,
  2018.

\bibitem{gimenez2011study}
E.~Gimenez, R.~Ballabriga, M.~Campbell, I.~Horswell, X.~Llopart, J.~Marchal,
  K.~Sawhney, N.~Tartoni, and D.~Turecek, ``Study of charge-sharing in medipix3
  using a micro-focused synchrotron beam,'' {\em Journal of Instrumentation},
  vol.~6, no.~01, p.~C01031, 2011.

\bibitem{van2016fast}
W.~Van~Aarle, W.~J. Palenstijn, J.~Cant, E.~Janssens, F.~Bleichrodt,
  A.~Dabravolski, J.~De~Beenhouwer, K.~J. Batenburg, and J.~Sijbers, ``Fast and
  flexible x-ray tomography using the astra toolbox,'' {\em Optics express},
  vol.~24, no.~22, pp.~25129--25147, 2016.

\bibitem{van2015astra}
W.~Van~Aarle, W.~J. Palenstijn, J.~De~Beenhouwer, T.~Altantzis, S.~Bals, K.~J.
  Batenburg, and J.~Sijbers, ``The astra toolbox: A platform for advanced
  algorithm development in electron tomography,'' {\em Ultramicroscopy},
  vol.~157, pp.~35--47, 2015.

\bibitem{reynolds2009gaussian}
D.~A. Reynolds {\em et~al.}, ``Gaussian mixture models.,'' {\em Encyclopedia of
  biometrics}, vol.~741, no.~659-663, 2009.

\bibitem{pan2002maximum}
J.-X. Pan and K.-T. Fang, ``Maximum likelihood estimation,'' in {\em Growth
  curve models and statistical diagnostics}, pp.~77--158, Springer, 2002.

\bibitem{scikit-learn}
F.~Pedregosa, G.~Varoquaux, A.~Gramfort, V.~Michel, B.~Thirion, O.~Grisel,
  M.~Blondel, P.~Prettenhofer, R.~Weiss, V.~Dubourg, J.~Vanderplas, A.~Passos,
  D.~Cournapeau, M.~Brucher, M.~Perrot, and E.~Duchesnay, ``Scikit-learn:
  Machine learning in {P}ython,'' {\em Journal of Machine Learning Research},
  vol.~12, pp.~2825--2830, 2011.

\bibitem{le2011least}
H.~Q. Le and S.~Molloi, ``Least squares parameter estimation methods for
  material decomposition with energy discriminating detectors,'' {\em Medical
  physics}, vol.~38, no.~1, pp.~245--255, 2011.

\bibitem{lee2014quantitative}
S.~Lee, Y.-N. Choi, and H.-J. Kim, ``Quantitative material decomposition using
  spectral computed tomography with an energy-resolved photon-counting
  detector,'' {\em Physics in Medicine \& Biology}, vol.~59, no.~18, p.~5457,
  2014.

\bibitem{5610726}
M.~Das, H.~C. Gifford, J.~M. O'Connor, and S.~J. Glick, ``Penalized maximum
  likelihood reconstruction for improved microcalcification detection in breast
  tomosynthesis,'' {\em IEEE Transactions on Medical Imaging}, vol.~30, no.~4,
  pp.~904--914, 2011.

\end{thebibliography}

\end{document}